\documentclass[prl,aps,twocolumn,showpacs,floatfix]{revtex4}

\usepackage{amsmath,amssymb}
\usepackage{graphicx}
\usepackage{version}
\DeclareGraphicsExtensions{.eps,.ps}

\newcommand{\bk}{{\bf k}}
\newcommand{\bK}{{\bf K}}

\newcommand{\bq}{{\bf q}}

\newcommand{\Max}{{\mathrm{Max}}}

\newcommand{\beqa}{\begin{eqnarray}}
\newcommand{\eeqa}{\end{eqnarray}}

\begin{document}


\title
{Excitonic condensation of massless fermions in graphene bilayers}
\author{C.-H. Zhang}
\affiliation{Department of Physics, Indiana University-Purdue
University Indianapolis (IUPUI), Indianapolis, Indiana 46202, USA}
\author{Yogesh N. Joglekar}
\affiliation{Department of Physics, Indiana University-Purdue
University Indianapolis (IUPUI), Indianapolis, Indiana 46202, USA}
\date{\today}

\begin{abstract}
Graphene, a single sheet of graphite with honeycomb lattice structure, has 
massless carriers with tunable density and polarity. We investigate the ground 
state phase diagram of two graphene sheets (embedded in a dielectric) 
separated by distance $d$ where the top layer has electrons and the bottom 
layer has holes, using mean-field theory. We find that a uniform excitonic 
condensate occurs over a large range of carrier densities and is weakly 
dependent on the relative orientation of the two sheets. We obtain the 
excitonic gap, 
quasiparticle energy and the density of states. We show that both, 
the condensate phase stiffness and the mass of the excitons, with 
massless particles as constituents, vary as the square-root of the carrier 
density, and predict that the condensate will not undergo Wigner 
crystallization.
\end{abstract}

\maketitle


\noindent{\it Introduction:} 
Over the past three years, graphene has emerged as the unique candidate that 
provides a realization of two-dimensional massless fermions whose carrier 
density and polarity are tunable by an external gate 
voltage~\cite{Novoselov05}. Subsequent experimental and theoretical 
investigations have led to a thorough re-examination of some of the 
properties of linearly dispersing massless fermions~\cite{riseofg,kyang}. 
The truly two dimensional (2D) nature of graphene permits control and 
observation of local carrier density and 
properties~\cite{markus,levitov,yacobi}. In graphene bilayers, the 
ability to change the carrier polarity of an individual layer implies 
that the interlayer Coulomb interaction can be tuned from repulsive to 
attractive. This raises the possibility of formation of electron-hole 
bound states or indirect excitons, albeit with massless fermions as its 
constituents. Properties of such bound states of 
{\it massless particles} are an open question; the only other example, 
to our knowledge, is the proposed color superconductivity in dense quark 
matter~\cite{alford}. Graphene bilayers provide an ideal and unique 
candidate for straightforward experimental investigations of such phenomena. 

A uniform Bose-Einstein condensate of excitons in electron-hole bilayers 
occurs when the interlayer distance is comparable to the distance between 
the particles within each layer~\cite{snokebook,Littlewood96}. These 
excitons have mass $m_{ex}=m_e+m_h$ where $m_e$ ($m_h$) is the band mass 
of the electron (hole). At high densities, dipolar repulsion between 
the excitons can lead to a condensate ground state with broken translational 
symmetry: a supersolid~\cite{Joglekar06}. Biased bilayer quantum Hall 
systems near total filling factor one have shown uniform excitonic 
condensation~\cite{ja}. In this case, the 
exciton mass is determined solely by interlayer Coulomb interaction and is 
independent of the bias voltage~\cite{kyangdipolar,biasedbilayers}. These 
observations raise the questions: What is the mass of an exciton with massless 
constituents? Will such an excitonic condensate lead to a supersolid if 
the dipolar repulsion between such excitons (with a nonzero mass) is 
increased? 

In this paper, we investigate the excitonic condensation in two graphene 
sheets embedded in a dielectric and separated by a distance $d\gg a$ 
($a$=1.4\AA\, is the honeycomb lattice size) so that the tunneling between the 
layers is negligible, but interlayer Coulomb interaction is not. The layers 
have opposite polarity and equal density of carriers $n_{2D}$. We remind the 
Reader that in graphene, in the continuum limit, the length-scale $1/k_F$ 
and the energy-scale $E_F$ are both set by the density of carriers $n_{2D}$ 
($k_F=\sqrt{\pi n_{2D}}$ is the Fermi momentum, $E_F=\hbar v_G k_F$ is the 
Fermi energy, and $v_G\sim c/300$ is the speed of massless carriers). 
Therefore, the ground state phase diagram depends only on one dimensionless 
parameter $k_Fd$. This is markedly different from conventional bilayer 
systems parameterized by ($d/a_B,r_s$) where $a_B$ is the band Bohr radius and 
$r_s=1/\sqrt{\pi a_B^2 n_{2D}}$~\cite{Littlewood96}, as well as biased bilayer 
quantum Hall systems, parameterized by ($d/l_B,\Delta\nu$) where $l_B$ is the 
magnetic length and $\Delta\nu$ is the filling factor 
imbalance~\cite{hanna,biasedbilayers,tutuc,biasedqhe}. 

We use the mean-field theory to obtain the ground-state phase diagram 
as a function of $k_Fd$. We find that 
a) excitonic condensation occurs at all densities as long as $k_Fd\sim 1$. 
b) the condensate properties are 
weakly sensitive to the relative orientation of the two sheets (stacking). 
c) the superfluid phase stiffness $\rho_s$ and the exciton mass have a 
$\sqrt{n_{2D}}$ dependence. d) the excitonic condensate does not undergo 
Wigner crystallization in spite of dipolar repulsion between excitons with 
a nonzero mass. 
 
The plan of the paper is as follows. In the next section, we present the 
mean-field Hamiltonian~\cite{negele} and briefly sketch the outline of 
our calculations. In the subsequent section, we show the results for the 
excitonic gap $\Delta_\bk$, the quasiparticle energy $E_\bk$, and the 
quasiparticle density of states $D(E)$. We discuss the density dependence 
of the superfluid stiffness $\rho_s$ and the mass of the excitons. In the last 
section, we show that these results are equivalent to absence of Wigner 
crystallization, and mention the implications of our results to experiments.  


\noindent{\it Mean-field Model:} We consider two graphene sheets embedded 
in a dielectric separated by distance $d$ with chemical potentials in the 
two layers 
adjusted so that the top layer (denoted by pseudospin 
$\tau=+1$) has electrons and the bottom layer (denoted by pseudospin 
$\tau=-1$) has holes with the same density. We 
consider two stackings: the Bernal stacking that occurs naturally in 
graphite, and the hexagonal stacking in which each sublattice 
($A$ and $B$) in one layer is on top of the corresponding sublattice in the 
other layer. Since the Hamiltonian in the continuum description is SU(4) 
symmetric in the spin and valley indices, we ignore those indices for 
simplicity. In the continuum limit, the single-particle Hamiltonian for 
carriers in layer $\tau$ is~\cite{Semenoff84} 
\begin{align}
\label{eq:h0}
\hat{H}_0=\Sigma_{\bk\alpha} (\alpha\hbar v_Gk)c^{\dagger}_{\bk\alpha\tau}
c_{\bk\alpha\tau} 
\end{align}
where $\bk$ is the momentum measured from the $\bK$-point and $\alpha=\pm$ 
denote the conduction and valance bands that result from diagonalizing the 
Hamiltonian in the sublattice-basis. $c^{\dagger}_{\bk\alpha\tau}$ 
($c_{\bk\alpha\tau}$) is creation (annihilation) operator for an electron 
in band $\alpha$ in layer $\tau$ with momentum $\bk$. We point out that for 
the hexagonal stacking, $c^{\dagger}_{\bk\alpha\tau}
=[c^{\dagger}_{\bk A\tau}+\alpha e^{-i\theta_\bk}c^{\dagger}_{\bk B\tau}]
/\sqrt{2}$ is independent of the layer index $\tau$. For the Bernal 
stacking, the creation operators in the two layers are related by 
{\it complex} conjugation, 
$c^{\dagger}_{\bk\alpha\tau}=[c^{\dagger}_{\bk A\tau}+\alpha 
e^{-i\tau\theta_\bk}c^{\dagger}_{\bk B\tau}]/\sqrt{2}$ where 
$\theta_\bk=\tan^{-1}(k_y/k_x)$. 
The interaction Hamiltonian consists of intralayer Coulomb 
repulsion $V_A(\bq)=2\pi e^2/\epsilon q$ and interlayer Coulomb attraction 
$V_E(\bq)=-V_A(\bq)\exp(-qd)$. ($\epsilon$ is the dielectric constant). 
Using standard mean-field techniques~\cite{negele}, we obtain the 
following mean-field Hamiltonian  
\begin{align}
\label{eq:HF}
\hat{H}=\sum_{\bk}
\left[\begin{array}{cc} e^{\dagger}_\bk & h_{-\bk}\end{array}\right]
\left[\begin{array}{cc} 
\epsilon_\bk-\mu & \Delta_\bk \\ 
\Delta^*_\bk & -\epsilon_\bk+\mu 
\end{array}\right]
\left[\begin{array}{c} e_{\bk}\\ h^{\dagger}_{-\bk}\end{array}\right].
\end{align}
Here $e^{\dagger}_{\bk}=c^{\dagger}_{\bk ++}$ creates an electron in the 
conduction band ($\alpha=+$) in the top layer ($\tau=+1$) and 
$h^{\dagger}_{-\bk}=c_{\bk --}$ creates a {\it hole} in the valance band 
($\alpha=-$) in the bottom layer ($\tau=-$). The term $\epsilon_\bk$ 
contains single-particle energy, capacitive Hartree self-energy and the 
intralayer exchange self-energy. The off-diagonal term $\Delta_\bk$ is 
proportional to the excitonic condensate order parameter 
$\langle h_{-\bk}e_{\bk}\rangle$. The eigenvalues 
of the mean-field Hamiltonian are given by $\pm E_\bk=\pm
\sqrt{(\epsilon_\bk-\mu)^2+\Delta^2_\bk}$. We consider mean-field 
states with a real $\Delta_\bk=\Delta^{*}_\bk$, and spatially uniform 
density. It is straightforward to diagonalize the Hamiltonian and 
obtain the mean-field equations~\cite{Littlewood96}
\begin{eqnarray}
\label{eq:epsilon}
\epsilon_{\bk}& = & \hbar v_G k+\frac{e^2 n_{2D}}{C}-
\frac{1}{2}\int_{\bk'} V_A(\bk-\bk')\left[1-\frac{\xi_\bk'}{E_\bk'}
\right]\\
\label{eq:delta}
\Delta_{\bk}& = & -\frac{1}{2}\int_{\bk'}V_E(\bk-\bk')f(\theta_{\bk,\bk'})
\frac{\Delta_{\bk'}}{E_\bk'} 
\end{eqnarray}
where $\xi_\bk=\epsilon_\bk-\mu$, 
$C=\epsilon/2\pi d$ is the capacitance per unit area, and 
$\theta_{\bk,\bk'}=\theta_\bk-\theta_{\bk'}$. The form factor for the two 
stackings are  
\begin{align}
f(\theta_{\bk,\bk'})=\left\{\begin{array}{cc}
(1+\cos\theta_{\bk,\bk'}) & \mathrm{Hexagonal}\\ 
\cos\theta_{\bk,\bk'}(1+\cos\theta_{\bk,\bk'})& \mathrm{Bernal} 
\end{array}\right..
\end{align}
We point out that the self-energy in Eq.(\ref{eq:epsilon}) takes into 
account both intrinsic and extrinsic contributions that cancel the 
$\cos\theta_{\bk,\bk'}$-dependent terms in the form 
factor and make the results independent of the ultra-violet 
cutoff~\cite{intrinsic,allan_vG}. Therefore the 
intra-layer self-energy in Eq.(\ref{eq:epsilon}) is the same as that 
for a conventional system~\cite{intrinsic,allan_vG}. 
The chemical potential $\mu$ is determined by the carrier density that 
takes into account the four-fold spin and valley degeneracy 
\begin{align}
\label{eq:mu}
n_{2D}=4\int_{\bk}\left[1-\frac{\xi_\bk}{E_\bk}\right]. 
\end{align}
It is straightforward to derive similar equations for a conventional 
electron-hole system~\cite{Littlewood96}. 
They are obtained by changing the single-particle dispersion to a quadratic 
and replacing the form factor $f(\theta_{\bk,\bk'})$ by a constant $f=2$. 
We solve Eqs. (\ref{eq:epsilon}), (\ref{eq:delta}) and (\ref{eq:mu}) 
iteratively to obtain self-consistent results. 


\begin{figure}[thbf]
\includegraphics[width=1\columnwidth]{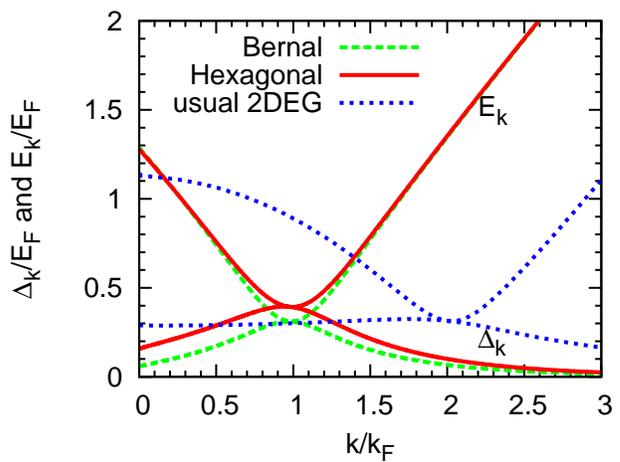}
\vspace{-8mm} 
\caption{(Color online) Excitonic gap $\Delta_\bk$ and the 
quasiparticle energy $E_\bk$ in graphene bilayer for Bernal (green dashed) 
and hexagonal (red solid) stacking with $k_Fd=1$. The quasiparticle 
spectrum $E_\bk$ becomes linear with a renormalized 
velocity $\tilde{v}_G>v_G$ for large $k\gg k_F$. The dotted blue curves 
show corresponding results for an electron-hole system at $r_s=2.7$ and 
$k_Fd=1$ when plotted using relevant (atomic) unit for 
energy~\cite{Littlewood96}.} 
\label{fig:deltaek}
\end{figure}

\noindent {\it Results:} Figure~\ref{fig:deltaek} shows the excitonic gap 
$\Delta_\bk$ and the quasiparticle energy $E_\bk$ for the Bernal (green 
dashed) and the hexagonal (red solid) stacking. The excitonic gap 
$\Delta_\bk$ is maximum at the Fermi momentum $k_F$  where the quasiparticle 
energy $E_\bk$ is minimum. Since 
the electron-hole Coulomb interaction is always attractive, the excitonic 
condensate order parameter is nonzero down to the bottom of the Fermi sea, 
$\Delta_{\bk=0}\neq 0$. Our results 
predict that the hexagonal-stacked system will have a larger excitonic gap 
than the Bernal-stacked system.The quasiparticle 
energy $E_\bk$ becomes linear at large $k\gg k_F$, since the constituent 
particles of the exciton have a linear dispersion. The 
speed of these quasiparticles is increased due to intralayer exchange 
self-energy~\cite{allan_vG,ds,roldan} although the increase is modest, 
$\sim 10\%$. Corresponding results for a conventional electron-hole system 
(blue dotted) are also shown in Fig.~\ref{fig:deltaek}.

\begin{figure}[thbf]
\includegraphics[width=1\columnwidth]{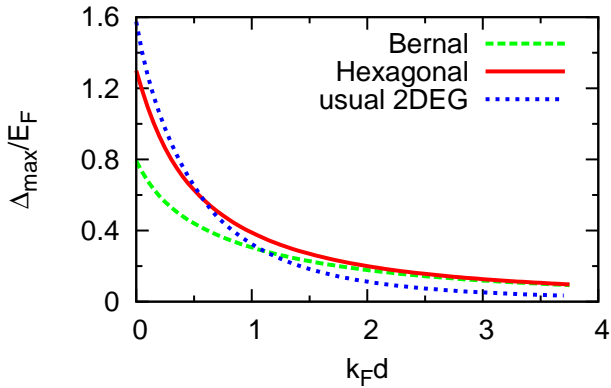}
\vspace{-8mm} 
\caption{(Color online) Dependence of the graphene bilayer excitonic gap 
$\Delta_m=\Max(\Delta_\bk)$ on interlayer distance $d$ for Bernal (green 
dashed) and hexagonal (red solid) stacking. This gap can be tuned by 
changing $n_{2D}$ for a given sample. Corresponding result for a conventional 
system at $r_s=2.7$ is shown in dotted blue.} 
\label{fig:delta_vs_d}
\end{figure}
Figure~\ref{fig:delta_vs_d} shows the dependence of the maximum excitonic 
gap $\Delta_{m}$ on $k_Fd$. We find that $\Delta_m$ is weakly dependent on 
the stacking and decays rapidly when $k_Fd\gg 1$. {\it This result 
implies that the excitonic condensation is a robust phenomenon that will not 
require precise alignment of the two graphene sheets when they are being 
embedded in a dielectric}. With typical graphene carrier densities 
$n_{2D}\sim 10^{12}$/cm$^2$ and $d\sim 100$\AA\, or $k_Fd\sim 1$, the 
excitonic gap is appreciable, $\Delta_m\sim 30$ meV. 

\begin{figure}[thbf]
\includegraphics[width=1\columnwidth]{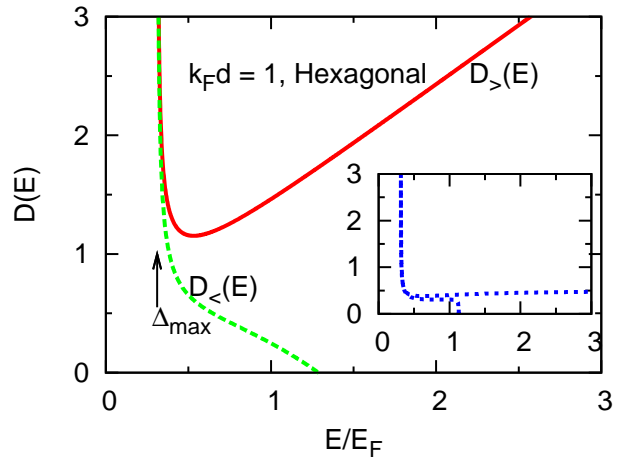}
\vspace{-8mm}
\caption{(Color online)Quasiparticle density of states contributions 
$D_<(E)$ (green dashed) associated with states with $k\leq k_F$, and $D_>(E)$ 
(red solid) associated with states $k\geq k_F$. These results are for 
hexagonal stacked graphene bilayers with $k_Fd=1$. Both diverge at 
$E=\Delta_m$, as expected. The total density of states $D=D_<+D_>$ can be 
probed by differential conductance for tunneling from a metal into the 
condensate. The inset shows corresponding results for an electron-hole 
system at $r_s=2.7$ and $k_Fd=1$. All results are expressed in their 
respective units.}
\label{fig:dos}
\end{figure}
A direct probe of the excitonic gap is the quasiparticle density of states. 
For graphene with no interactions, the density of states is linear, 
$D_0(E)=2E/\pi\hbar^2 v_G^2$. In the excitonic condensate phase, for 
intermediate energies $\Delta_m\leq E\leq E_{\bk=0}$ there are two rings 
in the phase-space consistent with that energy: one with $k<k_F$ and the other 
with $k>k_F$. Therefore the quasiparticle density of states is given by 
$D(E)=D_{<}(E)+D_{>}(E)$ where $D_{<}$ ($D_>$) denotes the density of states 
from respective rings. Figure~\ref{fig:dos} shows $D_<(E)$ (green dashed) 
and $D_>(E)$ (red solid); they are both zero for $E<\Delta_m$ and diverge 
at $\Delta_m$ as is expected. 
Note that $D_<(E)=0$ for $E>E_{\bk=0}$, since there are no states for $k<k_F$ 
with energies higher than $E_{\bk=0}$. The asymmetry in $D_<$ and $D_>$ for 
$E\gg\Delta_m$ is due to the linear dispersion of carriers and the nonzero 
electron-hole pairing that extends to the bottom of the Fermi 
sea, $\Delta_{\bk=0}\neq 0$~\cite{footnote}. The inset shows corresponding 
results a conventional system, where the density of states without 
interactions is constant, $D_0(E)=m/\pi\hbar^2$. 

Superfluidity of a uniform Bose-Einstein condensate is characterized by a 
non-zero phase stiffness $\rho_s$ that quantifies the energy of a 
condensate with a linearly winding phase, $E(Q)=\rho_s Q^2A/2$ where $A$ is 
the area of the sample and the phase of the condensate varies as $\Phi(x)=Qx$. 
For graphene, since $E_F$ is the sole energy scale (at zero temperature), it 
follows from dimensional analysis that phase stiffness must scale linearly 
with the Fermi energy, $\rho_s=g(k_Fd) E_F$ where $g(x)$ is a dimensionless 
function that satisfies $g\sim O(1)$~\cite{caveat2} when $0\leq x\lesssim 1$ 
and $g\rightarrow 0$ for $x\gg 1$. Hence, the phase stiffness is given by 
$\rho_s=g(k_Fd)\hbar v_G \sqrt{\pi n_{2D}}$. The condensate energy $E(Q)$ 
can also be expressed, in the particle-picture, as the kinetic energy of 
excitons that have condensed in a state with center-of-mass momentum
 $\hbar Q$. Thus, $E(Q)=N\hbar^2Q^2/2m_{ex}$ where $m_{ex}$ ($N$) is the mass 
(number) of condensed excitons~\cite{caveat3}. Equating the two expressions 
for energy implies $m_{ex}=n_{2D}\hbar^2/\rho_s\propto\sqrt{n_{2D}}$. 
Thus {\it we predict that the phase stiffness $\rho_s$ and the 
exciton mass will both vary as the square root of the carrier density}. 
We emphasize that these results are unique to graphene and, as we will 
show in the next section, are {\it equivalent} to the absence of 
excitonic Wigner crystallization in graphene 
bilayers~\cite{Chen91,Joglekar06}. 


\noindent{\it Discussion:} In this paper, we have investigated the 
properties of excitonic condensates in graphene bilayers. Our calculations 
predict that excitonic condensation will occur at all carrier densities as 
long as $k_F d\sim 1$, 
and that the strength of the condensate, as measured by the excitonic gap 
$\Delta_m$ is relatively insensitive to the stacking.  

The mean-field results presented in this paper are obtained at zero 
temperature $T$=0. (Finite temperature analysis gives a critical temperature 
$T_{MF}/E_F\sim 0.2$ or $T_{MF}\sim 20$ meV. This is an artifact of the 
mean-field approximation.) In two dimensions, the critical temperature $T_c$ 
for Bose-Einstein condensation is zero, but the superfluid properties 
survive for $T\leq T_{KT}$ where $T_{KT}$ is the Kosterlitz-Thouless 
transition temperature. Therefore, our results will be valid at nonzero 
temperature $T\ll T_{KT}$~\cite{allan_note}. A weak disorder will suppress the 
excitonic condensate order parameter and reduce the excitonic gap, an effect 
equivalent to increasing the value of $k_Fd$. Therefore, we have ignored 
the effects of a weak disorder potential. 

In our analysis, we have only considered excitonic condensation with uniform 
density. In conventional (quantum Hall electron-hole) bilayers, varying $d$ 
and $r_s$ ($\nu$) leads to excitonic condensates with lattice 
structure~\cite{Chen91,qhss07,Joglekar06}. The origin of the lattice 
structure is 
Wigner crystallization of carriers in an isolated layer at large $r_s$ 
(small $\nu$). Graphene does not undergo Wigner crystallization as its 
carrier density is changed~\cite{hari}. Therefore, we expect that the 
excitonic condensate in graphene bilayers remains uniform. Now we show that 
this result is equivalent to our predictions for density dependence of 
$\rho_s$ and $m_{ex}$. The quantum kinetic energy of an exciton, associated 
with localizing it within a distance $1/k_F$, is $K=\hbar^2k_F^2/2m_{ex}$. 
The potential energy due to the dipolar repulsion between them is 
$P=e^2d^2 k_F^3/\epsilon$. Hence their ratio is given by 
$P/K=e^2d^2 k_Fm_{ex}/\epsilon\hbar^2$. Wigner crystallization occurs when the 
ratio $P/K\gg 1$. This ratio will solely be a function of 
$k_Fd$ - no matter what the value of $d$ is - {\it if and only if} 
$m_{ex}\propto k_F=\sqrt{\pi n_{2D}}$. Therefore, results in the last section 
show that {\it the excitonic condensate in graphene will not undergo 
Wigner crystallization} in spite of the dipolar repulsion between excitons 
with a quadratic dispersion. This result, too, is 
unique to graphene and is markedly different from the behavior of dipolar 
excitonic condensates in conventional bilayers. It is 
interesting that the mass of these effective bosons has the same 
density dependence and order of magnitude as the cyclotron mass of fermionic 
carriers in graphene~\cite{Novoselov05}. 

The onset of excitonic condensation can be detected by a divergent 
interlayer drag~\cite{ga}. A uniform in-plane magnetic field 
$B_{||}$ between the two graphene sheets is expected to induce a 
(counterflow) supercurrent $J_d$ in such a condensate~\cite{ayp}, 
$J_d=2\rho_s e^2 dB_{||}/\hbar^2$. The phase stiffness $\rho_s$ and its 
density dependence can be directly obtained from experimental measurements 
of the counterflow supercurrent. The verification (or falsification) of our 
predictions, including the density dependence of $\rho_s$ and $m_{ex}$, 
will deepen our understanding of properties and condensation of excitons 
with massless fermions as constituent particles.  


\noindent{\it Acknowledgments:} 
This work was supported by the Research Support Funds Grant at IUPUI. After 
this work was completed, we became aware of a recent related 
work~\cite{allan_note}. 

\bibliographystyle{apsrev}
\bibliography{reference2}

\begin{thebibliography}{31}
\bibitem[{foo()}]{footnote}
\bibinfo{note}{In $N$-dimensions, fermions with dispersion $\epsilon_k\sim
  k^{\alpha}$ and excitonic gap $\Delta$ will have
  $D_{<(>)}(E)=D_0[\mu\mp\sqrt{E^2-\Delta^2}]^{N/\alpha-1}$ for $\Delta\leq
  E\leq \sqrt{\mu^2+\Delta^2}$, as $k$ reduces from $k_F$ to 0. Thus, for
  $N=\alpha$, $D_<(E)=D_>(E)=D_0$ where $D_0=E/\sqrt{E^2-\Delta^2}$.}
\bibitem[{cav({\natexlab{a}})}]{caveat2}
\bibinfo{note}{$T=0$ phase stiffness, calculated in Ref.~\cite{allan_note} via
  mean-field analysis reproduces the same functional form and implies that
  $g=1/4\pi$.}
\bibitem[{cav({\natexlab{b}})}]{caveat3}
\bibinfo{note}{In our model, exciton formation and condensation occur at the
  same $k_Fd$. Hence, the $T=0$ superfluid exciton density is the same as the
  carrier density $n_{2D}$.}
\end{thebibliography}


\end{document}